\documentclass[12pt, preprint]{aastex}
\usepackage{natbib}

\def\be{\begin{equation}}
\def\ee{\end{equation}}

\begin{document}
\title{Constraining vacuum gap models of pulsar radio emission using
the intensity modulation index} \author{Janusz
Gil\altaffilmark{1} \& Fredrick A. Jenet\altaffilmark{2}}
\altaffiltext{1}{Institute of Astronomy, University of Zielona
G\'ora, Lubuska 2, 65-265, Zielona G\'ora, Poland}
\altaffiltext{2}{California Institute of Technology, Jet Propulsion 
Laboratory\\ 4800 Oak Grove Drive,Pasadena, CA 91109}

\begin{abstract}

Recent observations suggest that the level of pulse-to-pulse intensity
modulation observed in a given radio pulsar may depend on its period
and period derivative. Such a ``modulation index relationship'' (MIR)
may be an important tool for determining the physical processes behind
the radio emission. In the context of sparking gap models, the exact
functional form of the MIR depends on the physical processes occurring
on the surface of the neutron star in the region known as the ``vacuum
gap.''  Several possible vacuum gap models are studied here in order
to determine the expected MIR for a given model. Current observations
are consistent with two of the four models studied: the curvature
radiation driven vacuum gap and the curvature radiation driven near
threshold vacuum gap (CR-NTVG). It is shown that the inverse Compton
scattering driven vacuum gap models are not supported by the current
data. Given that the current data best supports the CR-NTVG model, its
possible that all pulsars have strong ($\approx 10^{13}$ G) surface magnetic
fields.

\end{abstract}
\keywords{pulsars:general}

\section{INTRODUCTION}
The physical mechanism of the coherent pulsar radio emission has
remained elusive since their discovery over 30 years ago. The
observed high brightness temperatures together with enormous
amount of phenomenology exhibited makes these sources very
difficult to understand. However, it is generally accepted that
pulsar radio emission is generated within a dense
electron-positron plasma, created near the polar cap and flowing
along open magnetic field lines. The observed pulse-to-pulse
intensity modulation can arise from the time-dependent lateral
structure of this flow, probed once per pulsar period by the
observer's line of sight. \citet[][ RS75 henceforth]{rs75}
proposed a pulsar model in which the lateral structure was in the
form of localized spark discharges in a charge depleted region
near the stellar surface know as the ``vacuum gap.'' \citet[][
GS00 henceforth]{gs00} explored the RS75 model in an attempt to
relate the observed radio emission properties to a pulsar's period
$P$ and its derivative $\dot{P}$. They postulated that the polar
cap is populated as densely as possible with a number of sparks,
each having a characteristic size as well as separation from
adjacent sparks approximately equal to the height $h$ of the
vacuum gap acceleration region. This leads directly to the
so-called complexity parameter \be a=\frac{r_p}{h} \label{a}\ee
equal to the ratio of the polar cap radius to the characteristic
spark dimension. One can show that $a$ is approximately equal to
the maximum number of sparks across the polar cap and thus the
maximum number of sub-pulses that can appear in the pulse window.
Therefore, $a$ describes the complexity of single pulses and/or
mean profile (see GS00 for details).

The pulse-to-pulse intensity modulation can be quantified by the
phase resolved modulation index defined as \be
m(\phi)=\frac{\sqrt{<I_s(\phi)^2>-<I_s(\phi)>^2}}{<I_s(\phi)>}
,\label{mphi} \ee where $I_s(\phi)$ is the pulsar intensity at pulse
phase $\phi$ and the angle brackets represent averaging over the
ensemble of single pulses \citep[for details see][]{jap01}.  Within
the framework of sparking gap model the observed pulse-to-pulse
intensity modulation is due to the presence of a number sparks moving
over the polar cap either erratically or in an organized manner
(drifting). As the number of sparks increase, one expects to see less
pulse-to-pulse intensity modulation. Hence, the modulation index
(Eq.~[\ref{a}]) should be anticorrelated to the complexity parameter
(Eq.~[\ref{mphi}]), to the extent that at very high values of $a$
there should be no detectable intensity modulation (GS00).

In general, the complexity parameter (Eq.~[\ref{a}]) can be expressed
in the form $a\propto P^\beta\dot{P}^\gamma$. Since the modulation
index may be directly related to some unknown function of $a$, it is
useful to use the Spearman rank ordered correlation (SROC) technique
in order to determine if a correlation exists between $a$ and $m$. The
SROC technique is independent, up to a sign, of any monotonic function
applied to $m$ and/or $a$. In this case, only the ratio $\alpha =
\beta / \ \gamma$ is important. \citet[][ JG03 henceforth]{jg03a} used
the SROC technique to search for the proposed relationship, hereafter
referred to as a modulation index relationship (MIR), using a sample
of 12 pulsars. This preliminary investigation found evidence for a MIR
with $\alpha$ between $[-5,-2]$. The most significant value of $\alpha
= -2.7$. In general, a MIR will provide an important constraint on any
model of pulsar radio emission. Even its non-existence may be an
important piece of information. For the purposes of this letter, only
spark gap models will be considered. These models predict the
existence of a MIR together with a value of $\alpha$. The specific
value of $\alpha$ depends on the vacuum gap model in question. Thus,
an observationally determined range of $\alpha$ values will rule out various
vacuum gap models.

In the next section, four models of the vacuum gap acceleration region
are studied in order to determine the expected value of $\alpha$ for
each model. This letter is summarized in \S 3.

\section{MODELS OF VACUUM GAP ACCELERATION REGIONS}

All vacuum gap (VG) models assume that the free outflow of electrons
or ions from the polar cap surface is strongly inhibited, which leads
to formation of a charge depleted region just above the polar
cap. These regions are conventionally called the vacuum gap even if
the charge density is not exactly zero within it \citep[for review
see][]{gmg03}. The charge depletion
with respect to the corotational value \citep{gj69}
results in an extremely high
potential drop across the VG, which is discharged by photon induced
pair creation in the strong and curved surface magnetic field. The
details of this discharge depend on the strength and curvature of the
surface magnetic field and on the physical process that generate the
electron-positron pair plasma.

The analysis presented here will calculate the complexity parameter
for each VG model in question. In order to do this, both the vacuum
gap height, $h$, and the polar cap radius, $r_p$, need to be estimated. Both
of these quantities will depend on the structure of the magnetic field
near the surface of the neutron star.

Classically, the magnetic field is assumed to be dipolar. There is
a growing evidence that the actual surface magnetic field is
highly non-dipolar\citep[for review see][]{grg03,ug03}. The total
surface field, $B_s$, may be written as \be B_s=bB_d, \label{bees}
\ee where $b$ is a constant of proportionality which relates the
dipolar field component, $B_d$, to the total surface field
strength. For VG models that include the effects of non-dipolar
fields (i.e. $b \neq 1$), it will be useful to estimate $b$ in
terms of $P$ and $\dot{P}$. Two assumptions will be made in order
to do this. First, it will be assumed that the observationally
inferred surface magnetic field is actually a measure of the
surface dipole component of the field. This is a reasonable
assumption since the power radiated by the dipolar magnetic field
dominates the other multi-pole components. The surface dipolar
magnetic field strength, $B_d$, is estimated by \be B_d=6.4\times
10^{19}(P\dot{P})^{0.5} {\rm G}=2\times
10^{12}(P\dot{P}_{-15})^{0.5}{\rm G} , \label{bede} \ee where $P$
is the pulsar period and $\dot{P}$ is the period derivative with
respect to time \citep{st83}. The second assumption will be that
the total surface field, $B_s$, will be approximately equal to
$10^{13} G$  for all pulsars, independent of $P$ and
$\dot{P}$\citep[][ and references therein]{gm02}. With these two
assumption, $b$ may be written as \be
b=5B_{13}(P\dot{P}_{-15})^{-0.5} , \label{b} \ee where
$B_{13}=B_s/10^{13}$~G is of order unity.

The size of the polar cap is determined by those magnetic field lines
which connect to the interstellar medium magnetic field. For all the
VG models discussed below, the polar cap radius will be estimated
using \be r_p=b^{-0.5}R_6^{1.5}10^4P^{-0.5} {\rm cm}, \label{erpe} \ee
where $R_6$ is the neutron star radius in units of $10^6$ cm.
This is the classical dipolar field polar cap radius modified by a
factor of $b^{-0.5}$ due to magnetic flux conservation over the tube
of open magnetic field lines.

Four VG models will be discussed in the next four subsections. The
models are classified by the structure of the surface magnetic
field and the physics of the electron-positron pair plasma
production process. The prefix ``CR'' stands for curvature
radiation induced pair production \citep[][ RS75]{e66} while the
prefix ``ICS'' stands for resonant inverse Compton scattering
induced pair production\citep[][ and references therein]{zetal97}.
The suffix ``VG'' is used for models that assume $b=1$ (i.e. a
pure dipolar field) while the suffix ``NTVG'' is for ``near
threshold vacuum gap'' models that assume $b$ is given by Equation
\ref{b}. The results of the next four subsections are summarized
in Table \ref{table1} which gives $a$ and $\alpha$ for each model.

\begin{deluxetable}{llcc}
\tablecaption{\label{table1}Summary of Vacuum Gap models}
\tabletypesize{\scriptsize} 
\tablehead{\colhead{Model} & \colhead{Complexity Parameter ($a$)}
& \colhead{$\alpha$} & \colhead{References}} \startdata
CR-VG    & $2.9{\cal R}_6^{-2/7} R_6^{3/2}P^{-9/14}\dot{P}_{-15}^{2/7}$&-2.25&1,2\\
CR-NTVG  & $3.0{\cal R}_6^{-2/7} R_6^{3/2} B_{13}^{-1/14}P^{-19/28}\dot{P}_{-15}^{1/4}$&-2.71&3,4\\
ICS-VG   & $1.1{\cal R}_6^{-0.57} R_6^{3/2}P^{0.14}\dot{P}_{-15}^{0.79}$&0.18&3,5\\
ICS-NTVG & $4.5{\cal R}_6^{-0.57} R_6^{3/2}B_{13}^{0.5}P^{-0.39}\dot{P}_{-15}^{0.25}$&-1.56&3,4\\
\enddata

\tablerefs{(1) \citet{rs75}, (2) \citet{gs00}, (3) \citet{gm01},
(4) \citet{gm02}, (5) \citet{zetal97}}

\end{deluxetable}

\subsection{CR-VG model}

The model of RS75 is the prototype of a curvature radiation induced
vacuum gap. The gap height is determined by the condition $h=l_{ph}$,
where $l_{ph}$ is the mean free path for a photon to collide with the
strong magnetic field and form an electron-positron pair. This model is
valid for a dipolar magnetic field with $B_d \leq 4.4\times
10^{12}~G$. The gap height in this model is $h=(3.5\times 10^3){\cal
R}_6^{2/7}P^{1/7}\dot{P}_{-15}^{-2/7}$~cm \citep[RS75, ][]{gm02},
where ${\cal R}_6\approx 1$ is the radius of curvature of field lines
in units of neutron star radius $R=10^6$~cm. The complexity parameter,
$a$, takes the form $a = 2.9{\cal R}_6^{-2/7} R_6^{3/2}P^{-9/14}\dot{P}_{-15}^{2/7}$ and $\alpha = -2.25$. This
value of $\alpha$ is currently in the range allowed by the analysis of
JG03.

\subsection{CR-NTVG model}

The CR-VG model described above has a fundamental problem: the binding
energy is much too small to prevent thermionic emission from the polar
cap surface \citep[e.g.][]{as91,um95,um96}. \citet{gm01} argued that
this problem can be solved by assuming that pulsars have an extremely
strong, non-dipolar surface magnetic field whose magnitude,$B_s$, is
close to about $10^{13} G$, independent of $P$ and $\dot{P}$. In
such a strong magnetic field ($B_s \geq 4.4\times 10^{12}~G$
\citep[e.g.][]{um95}), the curvature radiated photons will pair produce
near the kinematic threshold \citep[e.g.][]{dh83}. Hence, this type of
model is called a curvature radiation induced near threshold vacuum
gap (CR-NTVG) model. The gap height in this model is $h=(3\times
10^3){\cal R}_6^{2/7}b^{-3/7}P^{3/14}\dot{P}_{-15}^{-3/14}$~cm
\citep[][ GM02 hereafter]{gm02}. Using Equations \ref{a}, \ref{b}, and
\ref{erpe} one obtains $a = 3.0{\cal R}_6^{-2/7} R_6^{3/2}
B_{13}^{-1/14}P^{-19/28}\dot{P}_{-15}^{1/4}$. This gives $\alpha =
-2.71$, amazingly close to the maximum significant value obtained by
JG03.

\subsection{ICS-VG model}

In this model, the electron-positron pairs are produced by inverse
Compton scattered seed photons in a relatively low strength dipolar
magnetic field\citep[][ and references therein]{zetal97}. The gap
height in this model is $h=(8.8\times 10^3){\cal
R}_6^{0.57}P^{-0.64}\dot{P}^{-0.79}$~cm \citep{zhm00,gm01}. Thus,
the complexity parameter is $a=1.1{\cal R}_6^{-0.57}
R_6^{1.5}P^{0.14}\dot{P}_{-15}^{0.79}$ and $\alpha = 0.18$. This
value is well outside the range of $\alpha$ values supported by
the analysis of JG03.

\subsection{ICS-NTVG model}

This is a near-threshold version of the ICS-VG model, valid for
$B_s>4.4\times 10^{12}$~G. The gap height in this model is
$h=(5\times 10)^3{\cal R}_6^{0.57}b^{-1}P^{-0.36}\dot{P}_{-15}^{-0.5}$~cm (GM02). Using
equations \ref{a} , \ref{b}, and \ref{erpe} one obtains
 $a= 4.5{\cal R}_6^{-0.57} R_6^{1.5}B_{13}^{0.5}P^{-0.39}\dot{P}_{-15}^{0.25}$ and
$\alpha = -1.56$. This value of $\alpha$ is also outside the current range of supported values.

\section{SUMMARY AND DISCUSSION}

By analyzing the pulse-to-pulse intensity fluctuations of a set of 12
pulsars, JG03 found a significant correlation between the measured
intensity modulation index (Eq.~[\ref{mphi}]) and $a(\alpha) =
P^\alpha\dot{P}$ when $\alpha$ is within the range [-5,-2]. The most
significant correlation was found when $\alpha=-2.7$. JG03 concluded
that this modulation index relationship (MIR) is consistent with the
RS75 vacuum gap model which predicts $\alpha=-2.25$. The work
presented here calculates the expected values of $\alpha$ for four models
of the vacuum gap including the RS75 model. Assuming that future
observations confirm the results of JG03, the inverse Compton
scattering (ICS) driven gap models can be ruled out since they predict
values of $\alpha$ outside the allowed range. Note that JG03 performed
their analysis on ``core'' emission components only, hence the ICS
models are only ruled out for such components. The curvature radiation
driven gap models are consistent with the JG03 results. Moreover, we
demonstrate that the near threshold vacuum gap model of GM02
reproduces the observed value of the most likely exponent
$\alpha=-2.7$. Since this model requires large amplitude, non-dipolar,
surface magnetic fields, confirmation of a MIR with $\alpha=-2.7$
could be evidence for non-dipolar surface fields of order $10^{13}
G$. Such fields may be generated by small-scale turbulent dynamo
action just after the formation of the neutron star \citep{ug03} or by
Hall-effect processes occurring on the stellar surface \citep{grg03}.





\begin{acknowledgements}
Part of this research was performed at the Jet Propulsion
Laboratory, California Institute of Technology, under contract
with the National Aeronautics and Space Administration. GJ
acknowledges the support of the Polish State Committee for
scientific research under Grant 2 P03D 008 19. We thank E. Gil for
technical help.
\end{acknowledgements}

{}

\begin{thebibliography}{}
\bibitem[Abrahams \& Shapiro(1991)]{as91} Abrahams, A.M., \& Shapiro, S. L. 1991, \apj, 374, 652
\bibitem[Daugherty \& Harding(1983)]{dh83} Daugherty J., \&
Harding, A.K. 1983, \apj, 273, 761
\bibitem[Erber(1966)]{e66} Erber, T. 1966, Rev. Mod. Phys., 38, 626
\bibitem[Geppert, Rheinhardt, \& Gil(2003)]{grg03} Geppert, V., Rheinhardt, M., \& Gil, J. 2003, A\&A Letters, accepted (astro-ph/0311121)
\bibitem[Gil \& Sendyk(2000)]{gs00} Gil, J. \& Sendyk, M. 2000,
\apj, 541, 351 (GS00)
\bibitem[Gil \& Mitra(2001)]{gm01} Gil, J. \& Mitra, D. 2001, ApJ, 550,
383
\bibitem[Gil \& Melikidze(2002)]{gm02} Gil, J. \& Melikidze, G.I. 2002, ApJ, 577,
909 (GM02)
\bibitem[Gil, Melikidze \& Geppert(2003)]{gmg03} Gil, J.,
Melikidze, G.I., \& Geppert, U. 2003, A\&A, 407, 315 (GMG03)
\bibitem[Goldreich \& Julian(1969)]{gj69} Goldreich, P. \& Julian, H. 1969, \apj, 157,
869
\bibitem[Jenet, Anderson \& Prince(2001)]{jap01} Jenet, F.A., Anderson, S.B., \& Price, T.A. 2001, \apj, 546, 394
\bibitem[Jenet \& Gil(2003a)]{jg03a} Jenet, F.A. \& Gil, J. 2003a,
\apj, 596, L215 (JG03)
\bibitem[Jenet \& Gil(2003b)]{jg03b} Jenet, F.A. \& Gil, J. 2003b,
\apj, submitted
\bibitem[Ruderman \& Sutherland(1975)]{rs75} Ruderman, M. A., \& Sutherland, P. G. 1975, \apj, 196, 51
\bibitem[Shapiro \& Teukolsky(1983)]{st83} Shapiro, S.L., \& Teukolsky
S.A. 1983, Black Holes, White Dwarfs, and Neutron Stars: The
Physics of Compact Objects (New York: Wiley)
\bibitem[Urpin \& Gil(2003)]{ug03} Urpin, U., \& Gil, J. 2003,
A\&A, in press (astro-ph/0311180)
\bibitem[Usov \& Melrose(1995)]{um95} Usov, V.V., \& Melrose, D. B. 1995, Aust. J.
Phys. 48, 571
\bibitem[Usov \& Melrose(1996)]{um96} Usov, V.V., \& Melrose, D. B. 1996, \apj, 464,
306
\bibitem[Zhang et al.(1997)]{zetal97} Zhang, B., Qiao, G.J., Lin, W.D.,
et al. 1997, ApJ, 478, 313
\bibitem[Zhang, Harding \& Muslimov(2000)]{zhm00} Zhang, B., Harding, A., \&
Muslimov, A.G. 2000, ApJ, 531, L135
\end{thebibliography}
\end{document}